\documentclass[conference]{IEEEtran}
\IEEEoverridecommandlockouts

\usepackage{cite}
\usepackage{amsmath,amssymb,amsfonts}
\usepackage{algorithmic}
\usepackage{graphicx}
\usepackage{textcomp}
\def\BibTeX{{\rm B\kern-.05em{\sc i\kern-.025em b}\kern-.08em
    T\kern-.1667em\lower.7ex\hbox{E}\kern-.125emX}}
  
\begin{document}

\title{Comparison of Self-Aware and Organic Computing Systems
{\footnotesize}
}

\author{\IEEEauthorblockN{ Najma Gill}
\IEEEauthorblockA{\textit{Intelligent Systems} \\
\textit{University Of Passau}\\
Passau, Germany \\
}

}
\maketitle

\begin{abstract}
With increasing complexity and heterogeneity of computing devices, it has become crucial for system to be autonomous, adaptive to dynamic environment, robust, flexible, and having so called self-*properties. These autonomous systems are called organic computing(OC) systems. OC system was proposed as a solution to tackle complex systems. Design time decisions have been shifted to run time in highly complex and interconnected systems as it is very hard to consider all scenarios and their appropriate actions in advance. Consequently, Self-awareness becomes crucial for these adaptive autonomous systems. To cope with evolving environment and changing user needs, system need to have knowledge about itself and its surroundings. Literature review shows that for autonomous and intelligent systems, researchers are concerned about knowledge acquisition, representation and learning which is necessary for a system to adapt. This paper is written to compare self-awareness and organic computing by discussing their definitions, properties and architecture.
\end{abstract}


\section{Introduction}
Computing systems are becoming more complex with the passage of time. Number of computing devices per person has increased. In past, one person used to have one device which has changed dramatically from past few years. For example, one can have smart watch, multiple laptops or mobile devices from different vendors. Multiple devices and their diversity pose challenges for developers and managers. To deal with these challenges, it has become crucial for system to have autonomous behavior for adaptation. Organic computing has presented solution for complex computing systems. For an autonomous technical system to have advance adaptive behavior, self-awareness is necessary.
Many papers has been written on organic and self-aware computing systems. Although these systems are related with each other, there is no clear difference written between them. It is important to understand what is relationship between them; How they differ from each other. This paper describes these two systems in a way that a clear relationship can be better understood.
\newcommand{\RomanNumeralCaps}[1]
    {\MakeUppercase{\romannumeral #1}}
This paper has been organized as follows: section \RomanNumeralCaps 2 describes self-awareness (Definition, properties and architecture), section \RomanNumeralCaps 3 describes organic computing (Definition, properties and architecture), section \RomanNumeralCaps 4 compares self-aware and organic computing system, and last section \RomanNumeralCaps 5 presents conclusion.
\section{Self-aware computing system}
Autonomous adaptive systems have become focus of researcher's study these days. An autonomous system requires self-awareness for advance adaptive behavior. Recently, researchers have begun to understand the need for self-awareness for these technical systems \cite{ref1}. Term self-awareness has been adopted from biology and cognitive science.  Self-awareness has been proposed as a means for advance autonomous adaptive behavior for these complex systems \cite{ref3}. These systems works under constraints of user goals which are further divided into local goals. For managing trade-off between local and global goals, awareness of system itself and its environment are crucial \cite{ref1}. “Fine grained” knowledge representation along with online-learning framework provides basis for reliable and efficient self-adaptation. In other words, it can be said that self-awareness is necessary for self-adaptation and self-management \cite{ref2}, \cite{ref5}. In fact, literature review of autonomous and intelligent systems shows that researchers are concerned about knowledge acquisition, representation, agent learning and architecture \cite{ref1}.  Lewis et al. claims that increasing self-awareness can increase a system’s ability to manage complex trade-offs in changing conditions \cite{ref4}. Self-awareness also produces emergent behavior. 
According to Cox, being aware of itself is not just about having knowledge of self but also about how to use it to create goals.
Here is the working definition as given by Lewis at al. \cite{ref2}. \\
To be self-aware a node\footnote{node, agent, subsystem and system has been used interchangeably in this paper.} must:
\begin{itemize}
\item Possess information about its internal state (private self-awareness).
\item Possess sufficient knowledge of its environment to determine how it is perceived by other parts of the system (public self-awareness).
\end{itemize}
Optionally, it might also:
\begin{itemize}
\item Possess knowledge of its role or importance within the wider system.
\item Possess knowledge about the likely effect of potential future actions or decisions.
\item Possess historical knowledge.
\item Select what is relevant knowledge and what is not.
\end{itemize}
To manage itself, system needs considerable amount of knowledge about it-self \cite{ref1}, \cite{ref2} and its environment \cite{ref1}. Self-awareness is concerned about availability, collection and representation of knowledge about a system, by that system \cite{ref2}. This knowledge helps in reasoning and smart decision making for adaptive behavior. It keeps its history and a reward by monitoring its behavior to update one or more of its components, to achieve its goals. It is goal-oriented in that, it takes actions automatically to meet the given goals. Clear separation between self-awareness and self-expression has been made to help designers evaluate process possibilities \cite{ref1}, \cite{ref2}. Self-expression determines system's resulting actions based on analysis of this knowledge. In other words, it is a decision making component.
There are mainly two types of self-awareness \cite{ref2} i.e. private self-awareness and public-awareness. Private self-awareness is agent's internal knowledge only known to agent itself; no other node or agent knows about it, until it shares with them \cite{ref2}. Additionally, it is externally unobservable \cite{ref2}. Furthermore, it includes internal state, behavior, history \cite{ref2}, context, goals, values, objectives, and experiences gained from environment. 
Public self-awareness on the other hand, is agent's knowledge about its environment, its role, social relationship and impact of its own actions on other agents (in case of collective system) external to itself.
Self-aware node must have private (internal) and public (external) knowledge and should continue updating this data throughout its life-time (computational self-awareness) for adaptation. This is how, it can continue learning ways to improve performance.
\subsection{Characteristics}
According to Agarwal et al. there are five design properties a self-aware system should have.
\begin{itemize}
\item Introspective or self-awareness: System can observe and optimize its behavior.
\item Adaptive: It can adapt to dynamic environment.
\item Self-healing: It can correct its errors. 
\item Goal-oriented: It works under strict constraint of stakeholder's\footnote{The term stakeholder includes humans, developer, owner, users, administers and other systems \cite{ref7} } given goal.
\item Approximate: It can automatically choose level of precision.
\end{itemize}
\subsection{Architecture}
Self-awareness from psychology has inspired engineering self-aware computing systems. Lewis et al. emphasizes to enrich self-adaptive architecture with computational self-awareness\textemdash a new notion of self-awareness and corresponding self-expression. Computational self-awareness is a process or set of processes concerned not only about knowledge, but also with the ways that system can use to update that knowledge \cite{ref1}. For example, by using online learning it can update its knowledge base \cite{ref1}. Lewis et al. developed framework to identify potential benefits of increased self-awareness. Lewis et al. point of view is that psychology based architectural aspects are common to a number of self-aware systems. For example monitor-analyze-plan-execute (MAPE) loop which is often augmented with knowledge to create MAPE-K. This is an observer/controller and three layer architecture for adaptive software systems. However, Lewis et al. presented a framework which differs from previously proposed ones in two ways. 
\begin{itemize}
\item Their framework has been inspired from psychology. They presented reference architecture and derived architecture patterns which explicitly consider different self-awareness levels. 
\item Their framework does not consider self-awareness to be added as additional management or control layer. Instead, it allows engineers to consider the entire system and its environment while building self-aware capabilities
\end{itemize}
This paper describes only the reference architecture. Reader is requested to read \cite{ref11}. It describes eight design patterns given by Lewis et al., which are just the variations of the reference architecture. Figure 1 shows the summary of the reference architecture. It consists of self-aware nodes. This node is not necessarily a physical system but it can be any conceptual container for system being considered. Additionally, it is level of abstraction, where knowledge acquisition, knowledge representation, adaptation, learning and behavior determination occurs.  Autonomous agent, running thread, a physical machine or combination of any of these are examples of a node.
Complexity of a self-aware node or a system depends on five levels of self-awareness, proposed by Neisser. Capability of a node is determined by level or combination of these levels it has. 
\begin{itemize}
\item Stimulus–aware: A node is stimulus-aware if it has knowledge of event or stimulus and knows how to respond to it. It is a prerequisite for all other levels of self-awareness \cite{ref5}. Stimulus can be both public and private. In this awareness, system only knows about current stimuli and has no knowledge of past and future stimuli \cite{ref1}.

\item Interaction-aware: In this awareness, system can learns that its interaction with external phenomena, i.e. environment and other systems, originates from stimulus-awareness.  It is based on external phenomena therefore, it is public self-awareness. A system may also learn from internal interactions with itself therefore, it can also be private self-awareness \cite{ref1}.

\item Time-aware: A node is time-aware if it has information about its history or experience \cite{ref10} and possible future phenomena. It may involve using explicit memory, time series modeling or prediction. It can be public and private \cite{ref1}, \cite{ref5}

\item Goal-aware: A system is goal aware if it has information of its current goal, objective, preferences and constraints \cite{ref1}. It ensures that system has access to its goals, can reason about, and manipulates them.  It can also be public (external goals) and private (internal goals).  Public in case when goals are shared in collective system. \cite{ref1}

\item Meta-self-awareness: The most advanced level in which system is aware of its own self-awareness levels and their execution. It allows metacognitive processes to analyze cost and benefit of maintaining specific awareness level. Online learning occurs at meta-self-awareness where models of node’s own behavior are built and acted upon \cite{ref5}. This awareness allows system to adapt with respect to self-awareness levels. For example, changing algorithm to achieve certain level or by deciding whether specific level is needed or not at all. It is a form of private self-awareness as it considers only internal knowledge. However, system may change focus from one goal to another when there is a change in working environment or its internal state. It can also trade-off between different goals and feedback from environment and states with the help of meta-self-awareness \cite{ref1}. System monitors its behavior in terms of goals.
\end{itemize}
Self-awareness can be a property of autonomous system or it can also be collective. In case of collective, Self-awareness is distributed, like ant colonies and immune system, across elementary units of the system. These units keep system robust in case of disturbances. These smaller entities work together to give sense of collective self \cite{ref1}. Hence, they are aware of global state \cite{ref2}. Additionally, each system is self-aware at one or more levels of self-awareness. For example, smart camera networks where there is no global view but self-awareness capabilities of individual camera create collective self. 
Sensors are used to collect data. Private data is collected with the help of internal sensors. Internal sensor measures internal aspects for instance, temperature or battery sensors \cite{ref1}. On the other hand, public information is gathered using external sensors. Light sensors, camera and microphone are examples of external sensors \cite{ref1}. Additionally, sensors also observe internal and external actions taken by node. This data is pre-processed, analyzed and presented as knowledge base to self-expression component.   
Self-expression processes use this data to take appropriate actions with the help of actuators. Self-expressive system does not have direct access to design-time goal of a node however, it is responsibility of self-awareness process to represent goal information in a precise, meaningful, and efficient way with the help of utility function.  It uses different decision mechanisms for a given knowledge base \cite{ref1}. A node may have complex and context dependent goals but it may be aware its current goal and present it to self-expression. Self-aware systems range from simple to more complex. Cognitive radio devices, smart cameras and s-bot are few examples of such systems \cite{ref1}. 

\begin{figure}[htbp]
\centerline{
\includegraphics[width=0.85\columnwidth]{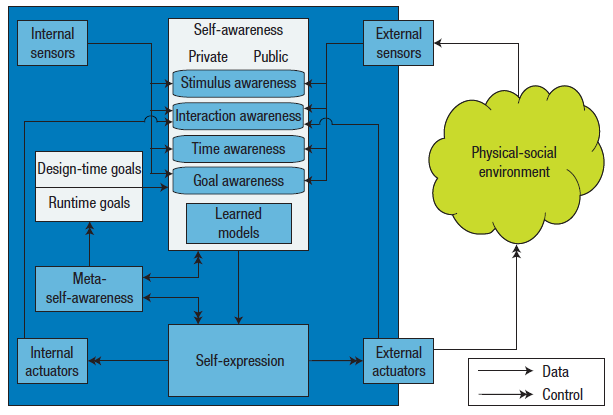}
}
\caption{ Reference architecture showing different self-awareness levels given by Lewis et al. [1]}
\label{fig1}
\end{figure}

\section{Organic Computing}
It is inspired from nature—distributed systems consisting of various autonomous systems. The term organic has been used with system because it is required of autonomous system to have properties similar to living creatures. It is also used for organization such as OC organizes autonomous systems in a large system \cite{ref6}. On the other hand, the term “self” corresponds to system’s autonomy \cite{ref6}. Design time decisions have been shifted to run time as in highly complex and interconnected systems, it is very hard to consider all scenarios and their appropriate actions in advance. System decides which resources it will use according to its environment, available resources, goals, and other constraints dynamically instead of traditional design method, where designer himself has to consider these constraints at design time. Furthermore, system can intelligently trade-off among available assets online. This is a great step forward toward complexity reduction where, user can give only high level goals to the system without concerning about how to do. Although, it has given relief to the designer but machine need to be equipped with enough intelligent methods and techniques to adapt \cite{ref5} to ever changing environment and user goals or needs \cite{ref8} in trustworthy way \cite{ref6}. System has been given more autonomy to reduce workload by stakeholders. On one hand, we want system to have more degree of freedom by allowing it to self-organize for adaptation. On the other hand, we do not want undesired and and anticipated behavior from system. Therefore, full autonomy is not desired, instead there is a need for controlled self-organized organic system so that human can control it \cite{ref9}.
An organic computing system is "a technical system which adapts dynamically to the current conditions of its environment. It will be self-organizing, self-configuring, self-healing, self-protecting, self-explaining, and context-aware" \cite[p.~25]{ref9}. It is necessary for OC system to survive in real environment \cite{ref7}. An autonomous system which stay robust against disturbances, adapts to dynamic environment and have self-* properties such as self-organization, self-healing, self-configuration, self-protection, self-optimization, is called OC system. It should adjust to human needs in trustworthy way and should provide explicit way for intervention in case of undesired behavior \cite{ref6}. 
According to Tomforde et al. self-organization and self-adaptation are just the means for making system robust. Robustness means resilient against internal or external disturbances or attacks. OC system consists of two parts\textemdash Productive part which satisfies the technical purpose and adaptive part responsible for creating organic capabilities in system \cite{ref6}. Machine learning techniques are used by OC system to react to unpredicted situations. Furthermore, such system do not waste valuable resources in finding optimal solution \cite{ref6}. Instead, it tries to react with good enough solution in real time with acceptable performance in presence of disturbances.
\subsection{Characteristics}
According to Tomforde et al. following are some of the properties of an organic system.
\begin{itemize}
\item System structure and behavior can be modified at run time.
\item Self-configuration: also called self-adaptation. According to higher level goals from user, system changes its parameters which results in change of behavior. 
\item Self-organization: It is related to change of structure of system. Some components can leave while others can join according to certain goal. 
\item Self-integration: It uses both self-configuration and self-organization. System autonomously decides its role and adapts its behavior and relation to other systems. 
\item Self-management: It involves both self-configuration and self-organization and other self-* properties.
\end{itemize}

Considering heterogeneity of modern day computing systems, generic architecture has been used so that they can communicate with each other easily. Therefore, Generic Observer/Controller \cite{ref6} architecture has been presented here. It is a layered and feedback based architecture, where higher layer shows higher abstraction level. Feedback mechanism allows system to continuously monitor it performance for adapting in dynamic environment. Adaptation means system continues learning from its previous actions and chooses best possible action according to situation at run time in accordance with its objective, resources and other constraints. Generic Observer/Controller has three main components. Figure 2 summarizes this architecture. 
\begin{figure}[htbp]
\centerline{
\includegraphics[width=0.85\columnwidth]{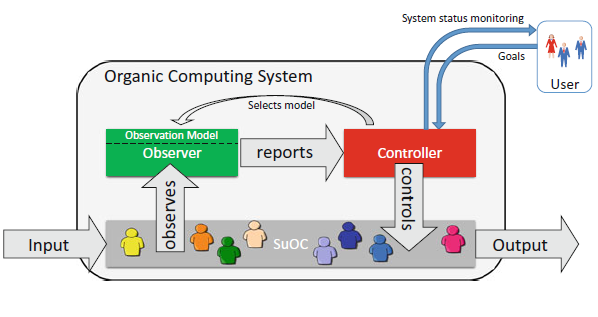}
}
\caption{ Detailed system design showing different layers. \cite{ref10}}
\label{fig2}
\end{figure} 

\subsubsection{System under Observation and Control (SuoC)} It is also called productive part.
It is encapsulated in lower layer 0 of architecture. Higher layers control SuoC by configuring its parameters. However, it is independent of upper layers; Therefore, it remains operational even when these layers fail. As OC system may consists of autonomous subsystems, SuoC can refer to single or group of systems.  According to Tomforde et al. SuoC has to meet basic requirements.
\begin{itemize}
\item Behavior of SuoC and its environment have to be observable.  
\item Performance of SuoC has to be measurable according to user given goal.
\item It has to have some parameters that can be changed at run time.
\end{itemize}

Layer 1 continues monitoring layer 0. The layer 1 and layer 2 plays a key role in learning of the system. Learning enables system to adapt. In this layer existing set of rules are followed due to safety reasons as untested solution is unacceptable. For any situation description from SuoC, situation is matched with available set of rules. Once it matches, specific action is performed in response otherwise similar most promising action is performed. Simultaneously, control is passed to layer 2 to generate new rules in case of unknown situations. Hence, system changes its parameters according to its environment. Additionally, action performed is also evaluated in layer 1. This layer is restricted to use given rules however, controller has option to select best strategy by learning. Layer 2 uses optimization heuristic to generate new rule after getting request from layer 1. The newly generated rules are then added to rule set of layer 1. Layer 3 serves as an interface for user and neighbouring nodes. Using this user can specify goals for the system.\cite{ref12}. Figure 3 show detailed view of these layers.
\begin{figure}[htbp]
\centerline{
\includegraphics[width=0.85\columnwidth]{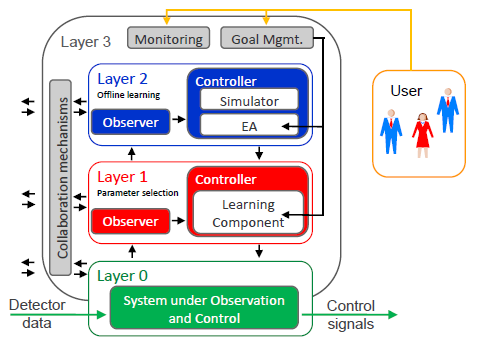}
}
\caption{ Detailed system design showing different layers. \cite{ref12}}
\label{fig3}
\end{figure}

\subsubsection{Observer}
It observes current state of SuoC. Subtasks of observation are monitoring, pre-processing, data analysis, prediction and aggregation \cite{ref8}. The attributes to be observed, are specified by observation model at run time. This observation model is constantly being optimized by controller according to goals and other constraints (e.g. resources). Observer creates situation description, which is later used by controller for decisions making. Monitoring of surroundings and system itself is done with the help of internal and external sensors.
\subsection{Controller}
Controller is the component, where user interacts with system and provides high level goals. Here user specifies what to do and not how to do. How-to-do is done by controller. It takes decisions itself based on situation description, without human intervention. Controller needs Sensor data, internal status of SuoC, utility function (user goal specific) and access to control interface of SuoC \cite[p.~174]{ref7} to take decisions. It might also have access to actuators of SuoC.
Controller is present in two layers: \cite{ref7}, \cite{ref8} layer 1 and layer 2.  Layer 1 is responsible for online learning where mapping of possible actions to known situations \cite{ref8} is done. After action execution and its evaluation, situation-action mapping gets updated. Layer 2 does offline learning and improves mapping of layer 1. It consists of adaptation and simulation module. Adaptation module creates additional mappings by using optimization or machine learning algorithms \cite{ref8}. Additionally, this module also relies on simulation model for new situation-action mapping. This new mapping is then added to previous mapping in layer 1. Actions are performed on actuators through SuoC or directly by controller.

\section{Comparison of Self-Aware and Organic Computing Systems}
As it has been explained in previous sections that OC system, is a technical system which adapts dynamically to its environment and have so called self-* properties. Neither fully autonomous nor fully controlled system is desired in organic computing. Rather, it is controlled self-organized system. On the other hand, self-awareness can be a property of an autonomous system, that allows it to realize advance adaptive behavior. Therefore, it is crucial for OC system to have self-awareness capability. System capable of self-awareness is called self-aware system. Self-aware system is aware of itself and its environment. 
It can be stated that it is self-awareness which makes OC system to adapt and learn through experience. Furthermore, knowledge gathering, analyzing, predicting, and expressing is also due to this property. Such a system keeps on learning throughout its life time. It is necessary for self-aware system to use its knowledge for better decision. Self-awareness helps to achieve self-management and advance adaptive properties. Technical autonomous system observes it environment using sensors and perform actions with help of actuators. Robustness is only present in OC system not in self-aware system.
 
\section{Conclusion}
This paper has explained self-aware and organic computing systems. For making a clear comparison it has described definition, properties and architecture of these two systems. It has also explained, how adaptation is realized in autonomous system, how self-awareness is required for OC to survive in real world,and how an autonomous system learns. 



\end{document}